\documentclass[%
 reprint,
 superscriptaddress,
 amsmath,amssymb,
 aps,longbibliography
prb,
]{revtex4-1}

\usepackage{graphicx}
\usepackage{dcolumn}
\usepackage{bm}
\usepackage{hyperref}
\usepackage[mathlines]{lineno}
\usepackage[bottom]{footmisc}
\usepackage[super]{nth}%
\usepackage{color}
\usepackage{hhline}

\begin{document}
\raggedbottom


\title{Sub-diffusive wave transport and weak localization transition in three-dimensional stealthy hyperuniform disordered systems}

\author{F. Sgrignuoli}
\affiliation{Department of Electrical and Computer Engineering \& Photonics Center, Boston University, 8 Saint Mary's Street, Boston, Massachusetts 02215, USA}
\author{S. Torquato}
\affiliation{Department of Chemistry, Department of Physics, Princeton Institute for the Science and Technology of Materials, and Program in Applied and Computational Mathematics, Princeton University, Princeton, New Jersey 08540, USA}
\author{L. Dal Negro}
\email{dalnegro@bu.edu}
\affiliation{Department of Electrical and Computer Engineering \& Photonics Center, Boston University, 8 Saint Mary's Street, Boston, Massachusetts 02215, USA}
\affiliation{Division of Material Science and Engineering, Boston University, 15 Saint Mary's Street, Brookline, Massachusetts 02446, USA}
\affiliation{Department of Physics, Boston University, 590 Commonwealth Avenue, Boston, Massachusetts 02215, USA}
\begin{abstract}
 The purpose of this work is to understand the fundamental connection between structural correlations and light localization in three-dimensional (3D) open scattering systems of finite size. We numerically investigate the transport of vector electromagnetic waves scattered by resonant electric dipoles spatially arranged in 3D space by stealthy hyperuniform disordered point patterns. Three-dimensional stealthy hyperuniform disordered systems (3D-SHDS) are engineered with different structural correlation properties determined by their degree of stealthiness $\chi$. Such fine control of exotic states of amorphous matter enables the systematic design of optical media that interpolate in a tunable fashion between uncorrelated random structures and crystalline materials. By solving the electromagnetic multiple scattering problem using Green's matrix spectral method, we establish a transport phase diagram that demonstrates a distinctive transition from a diffusive to a weak localization regime beyond a critical scattering density that depends on $\chi$. The transition is characterized by studying the Thouless number and the spectral statistics of the scattering resonances. In particular, by tuning the $\chi$ parameter, we demonstrate large spectral gaps and suppressed sub-radiant proximity resonances, facilitating light localization. Moreover, consistently with previous studies, our results show a region of the transport phase diagram where the investigated scattering systems become transparent. Our work provides a systematic description of the transport and weak localization properties of light in stealthy hyperuniform structures and motivates the engineering of novel photonic systems  with enhanced light-matter interactions for applications to both classical and quantum devices. 
 
\end{abstract}

\pacs{Valid PACS appear here}
\maketitle
\section{Introduction}\label{Introduction}
Understanding the subtle mechanisms behind the confinement of optical waves in complex media is essential to manipulate scattering and radiation phenomena. One of the simplest examples to achieve such control is introducing a defect in a periodic structure. Such a defect creates a localized state in the bandgap region whose confinement properties depend on the nature of the defect itself \cite{Joannopoulos}. For example, a point defect behaves like a micro-cavity, a line defect as a waveguide, and a planar defect like a perfect mirror \cite{Joannopoulos}. However, fabrication imperfections fundamentally limit the spatial extent of defect-localized modes in photonic bandgap structures \cite{Faggiani}, motivating the development of more robust alternatives. Controlling highly-localized optical states is also essential for quantum science and technology. For example, photonic environments capable of spatially confine light can enable the preparation of qubits that are well isolated from the environments reducing decoherence \cite{weiss2017quantum,branderhorst2008coherent}.

A promising approach to realize highly-confined optical modes relies on the design of strongly uncorrelated scattering materials where the propagation of light experiences coherent effects due to multiple interference phenomena. One of the most fascinating phenomena in the strong multiple scattering regime is the breakdown of the classical diffusion picture and the emergence of localized wave solutions \cite{LagendijkToday}. Initially proposed by P. W. Anderson in the context of electron waves transport \cite{Anderson}, disorder-induced localization of electromagnetic waves drives the development of disordered photonics \cite{WiersmaReview}. Since Anderson's Nobel-prize-winning discovery \cite{anderson1985question}, the study of how light propagates in uncorrelated materials unveiled a new and efficient method of trapping light leading to interesting and unexpected physical phenomena \cite{WiersmaReview,Yu}. In particular, disorder-engineered optical devices enable the control of the spectral, transport, and topological properties of light \cite{Yu}, offering new perspectives for both fundamental physics and technological applications \cite{WiersmaReview,ozawa2019topological}. For example, random lasers have been developed with both uniform \cite{Cao,DiederikLaser} and correlated disorder \cite{DalNegroScirep,ChenFractional,yoshimura2012secure,gaio2019nanophotonic}, leading to remarkable innovations in spectroscopy \cite{Redding,Boschetti} and optical imaging \cite{Sebbah,Bertolotti,Mosk}. Moreover, localized optical patterns promise to open new avenues for the encoding of both classical and quantum optical information \cite{leonetti2016secure,crespi2013anderson}. However, disorder-induced localization phenomenona depend critically on the dimensionality of the investigated systems \cite{LagendijkToday}. 
 
In fact, while light localization always occurs in one- and two-dimensional structures with sufficiently large size \cite{Chabanov,Anderson,RiboliALin2D,SgrignuoliPeenwheel}, this is not the case for three-dimensional (3D) disordered systems where Anderson localization has not been unambiguously demonstrated. Although several claims have been made \cite{WiersmaLocaliz,Storzer,SperlingAL}, the debate is still ongoing \cite{Beek2012,Sperling,Scheffold}. From the theoretical standpoint, it has been recently demonstrated that the localization of electromagnetic waves cannot be achieved in random ensembles of electric point scatterers, irrespective of their scattering strength \cite{SkipetrovPRL,Bellando,van2021longitudinal}. The near-field dipole-dipole coupling between vector scatterers was conjectured to be an important factor that prevents the onset of a delocalization-localization transition \cite{SkipetrovPRL,Bellando,van2021longitudinal,cobus2021transient,naraghi2016phase, SgrignuoliSubrandom}. Consequently, alternative strategies have been developed to reduce the detrimental effect of mediated dipole-dipole interactions. For instance, the application of strong magnetic fields in random media \cite{Skipetrov2015,Skipetrov2018}, the use of partially ordered media \cite{Skipetrov2020}, and complex aperiodic structures with \textit{ad hoc} structural properties \cite{SgrignuoliSubrandom,Haberko} have all been proposed as candidate systems for the demonstration of 3D light localization. In this context, disordered hyperuniform structures \cite{Torquato,TorquatoReview,Zachary} have attracted significant attention because they enable the design of unique materials that flexibly interpolate between the two extremes of ordered and disordered structures. These engineered disordered media resulted in novel optical phenomena in nanophotonics, such as the discovery of exotic amorphous materials with large, complete, photonic bandgaps \cite{Florescu}, the design of free-form waveguides \cite{Man}, and enhanced directional extraction of the incoherent emission from light-emitting diodes \cite{Gorsky}. Moreover, 3D photonic hyperuniform networks can be conveniently fabricated by silicon double inversion of polymer templates \cite{Muller}.
The development of a suitable platform for light localization provides opportunities to  enhance light-matter interactions in photonic materials with engineered disorder as well as novel quantum simulation protocols that are inherently more resilient to decoherence effects  \cite{BarredoScience,schreiber2011decoherence,wang2016single,viebahn2019matter,abouraddy2012anderson}. 

In this work, we study the transport and localization properties of electromagnetic waves scattered by resonant electric dipoles spatially arranged in 3D stealthy hyperuniform disordered configurations. These systems  suppress long-wavelength density fluctuations and the single-scattering of waves in a predefined spectral range around the origin of their reciprocal space \cite{Batten,Uche}. Motivated by the recent discovery that the optical transmittance of 3D hyperuniform systems can be strongly enhanced or suppressed depending on the frequency of the incident waves \cite{torquato2021nonlocal,Haberko}, we ask how structural correlations in 3D-SHDS more generally determine their transport and localization properties. A similar approach has been used to classify, both theoretically \cite{Leseur,Froufe} and experimentally \cite{Aubry}, the optical transport properties of two-dimensional SHDS, unveiling disorder-induced transparency, Anderson localization, and bandgap formation. In the present work, we extend this analysis to the more complex situation of electromagnetic vector waves scattered by a large number of 3D hyperuniform disordered point patterns characterized by different degrees of stealthiness $\chi$. To rigorously investigate wave localization in these 3D scattering open environments, we use the Green's matrix spectral method to study their complex scattering resonances systematically \cite{Lagendijk}. Moreover, we establish a characteristic transport phase diagram by computing the minimum value of the Thouless number \cite{Thouless} as a function of $\chi$ and the scattering strength of the investigated systems. Our results unveil the presence of a transition from a diffusive to sub-diffusive transport regime (weak localization) controlled by the stealthiness $\chi$ parameter and demonstrate the fundamental connection between hyperuniform structural correlations and light localization in 3D open scattering systems of finite size.      

\section{Characterization of disordered hyperuniform structures.}\label{Geometry}

Hyperuniform systems are structures with suppressed infinite-wavelength density fluctuations \cite{Torquato}. This large-scale property implies vanishing of the structure factor $S(\textbf{k})$ when $|\textbf{k}|\rightarrow0$ in the thermodynamic limit \cite{Torquato}. Because $S(\textbf{k})$ is proportional to the scattering intensity \cite{Bohren}, hyperuniformity results in the complete suppression of wave scattering as $|\textbf{k}|\rightarrow0$ \cite{TorquatoReview}. All crystals and quasicrystals are hyperuniform \cite{TorquatoReview}  as well as certain exotic amorphous structures \cite{Torquato,Batten,Uche,Zhangglass,Torquatorandom,SgrignuoliPeenwheel}. Hyperuniform disorder enables one to engineer materials \cite{WiersmaReview,Yu} with singular features, such as the control of the spectral (i.e., photonic bandgap) \cite{Man,Florescu} and momentum (i.e., directivity) \cite{Gorsky} responses of light. 
Moreover, stealthy hyperuniform structures are characterized by the fact that their structure factor vanishes over an entire domain $\Omega$ with size $K$, i.e., $S(\textbf{k})=0$ for all $|\textbf{k}|<K$ \cite{TorquatoReview,Batten,Uche}. 
Stealthy hyperuniformity is a stronger condition than  what achieved in standard hyperuniform media because single scattering events are prohibited for a large range of wavelengths. Therefore, stealthy hyperuniform systems provide unprecedented opportunities to engineer amorphous materials with tunable structural properties governed by their degree of stealthiness \cite{Torquato,Yu,TorquatoReview}. 

In this work, we have generated stealthy 3D disorder hyperuniform structures using the collective coordinate numerical optimization procedure to tailor the small-$k$ behavior of the structure factor \cite{Batten,Uche}. Specifically, this method is characterized by an initial uniform random distribution of $N$ points in a cubic simulation box of side length $L$. Then, the points are collectively moved, under periodic boundary conditions, to minimize a pair potential $\Phi$ within an exceedingly small numerical tolerance. In this way, the final structure factor is stealthy up to a finite cut-off wavenumber $K$. More details on this optimization procedure, as well as the definition of the pair potential $\Phi$, can be found in Refs.\,\cite{Batten,Uche,Torquato,TorquatoReview,Gorsky,torquato2015ensemble}

\begin{figure}[t!]
\centering
\includegraphics[width=\columnwidth]{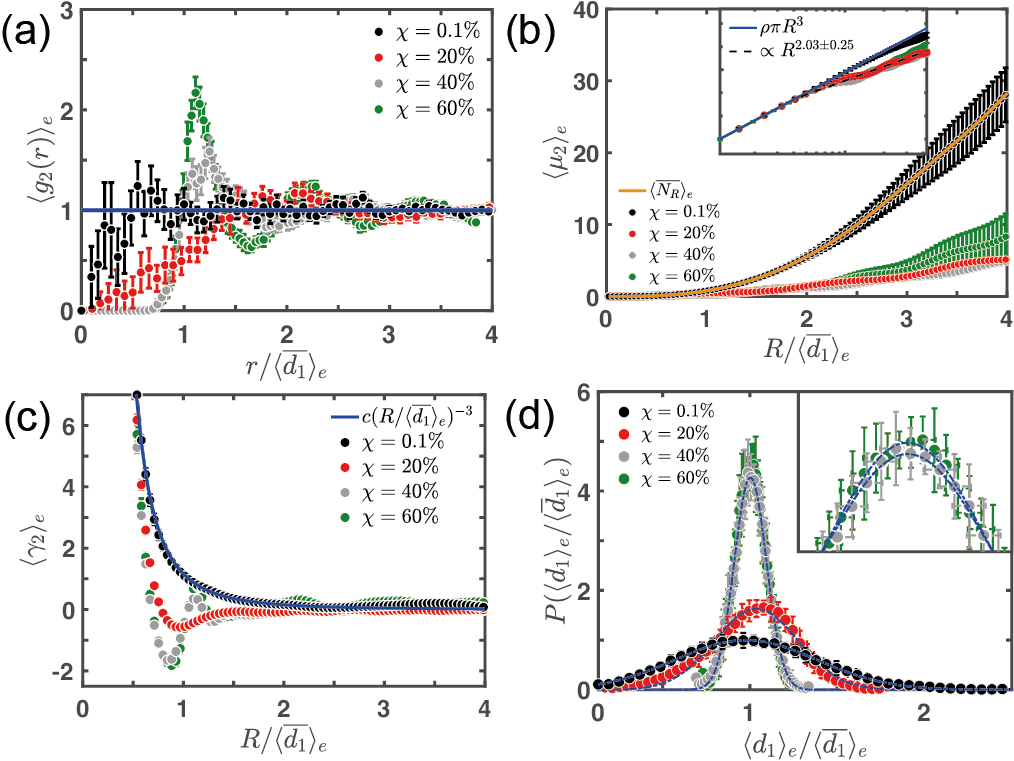}
\caption{Structural properties of 3D-SHDS composed by 1000 scattering elements and characterized by different $\chi$ values. In particular, $\chi$ is equals to $0.1\%$ (black lines), $20\%$ (red lines), $40\%$ (grey lines), and $60\%$ (green lines), respectively. Panel (a) shows the ensemble-averaged two-point correlation function $\langle g_2(r)\rangle_e$, while the scaling of the number variance of variance $\mu_2$ within a spherical observation window of radius $R$, averaged with respect to 10 different disordered realizations, is reported in panel (b). The inset of panel (b) shows the $\langle \mu_2\rangle_e$ data in a log-log scale. Moreover, the inset shows the prediction of the number of variance of uniform random point processes in the blue line. Panel (c) displays $\langle\gamma_2\rangle_e$ along with the analytical prediction related to uncorrelated Poisson processes in blue line. The probability density functions of the ensemble-averaged nearest spacing distances $\langle d_1\rangle_e$ normalized by $\langle\overline{d_1}\rangle_e$ is reported in (d). The error bars are the statistical errors associated with the average ensemble operation.}
\label{Fig1}
\end{figure}

In three spatial dimensions, an ensemble of N points represents a total of $3N$ degrees of freedom (DOF). The number of constrained points M(K) depends on the size $K$ of the reciprocal domain $\Omega$. The structural order in hyperuniform structures is controlled through the parameter $\chi$ that represents the degree of stealthiness, which  is defined as the ratio $M(K)/3N$ \cite{Uche}. For $\chi=0\%$, the system is uncorrelated, while for $\chi=100\%$, the system is a perfect crystal. The degree of short-range order in these structures increases with $\chi$, inducing a transition from a disordered ground state to a crystalline phase when $\chi>50\%$ \cite{Uche,torquato2015ensemble}. Moreover, increasing $\chi$ tends to increase the net repulsion of the potential $\Phi$. This feature is visible in Fig.\,(\ref{Fig1})\,(a), where we report the ensemble-averaged of the two-point correlation function $g_2(r)$ performed over ten different disorder realizations with $N=1000$ particles. This function describes how the particle's density varies as a function of distance, and it is proportional to the probability of finding two particles separated by a distance $r$ \cite{Illian}. In what follows, we will use the symbol $\langle\cdots\rangle_e$ to identify this ensemble-averaged operation. Representative examples of generated 3D-SHDS point patterns, with a tolerance in the minimization process of $10^{-12}$, are shown in the Appendix \ref{Appendice}. We have carefully checked that the tolerance threshold and different stealthy hyperunifomr optimization processes do not influence the main results of our paper, as discussed in the Appendix\,\ref{Appendice2}. While for $\chi=0.1\%$, the $\langle g_2(r)\rangle_e$ fluctuates around one (i.e., uncorrelated disorder), it develops an exclusion region when $\chi>20\%$ where $\langle g_2(r)\rangle_e$ is exactly zero for a domain near the origin. The extent of this exclusion region (i.e. $r/\langle \overline{d_1}\rangle\rightarrow 0$) increases with $\chi$. This feature is clearly visible in Fig.\,(\ref{Fig1}) by comparing the red curve (relative to $\chi=20\%$) with respect to the grey line (corresponding to $\chi=40\%$). The symbol $\langle\overline{d_1}\rangle_e$ corresponds to the mean value of the ensemble-averaged nearest spacing distances of the investigated structures. At $\chi=60\%$ (i.e., green curve), the peaks demonstrate crystallinity. This analysis shows that the investigated structures display the same features as the 3D stealthy hyperuniform point patterns already reported in literature\,\cite{Batten,Uche}.

Hyperuniformity can also be identified by analyzing the scaling of the fluctuations of the number of points $N_R$ contained within a spherical window of radius $R$ \cite{Torquato}. This scaling is quantified by the growth of the variance $\mu_2=\langle N_R^2\rangle-\langle N_R\rangle$ with respect to $R$ \cite{TorquatoReview,torquato2021strScaling}. If $\mu_2$ grows more slowly than the volume of the window in the large-$R$ limit, then the structure is hyperuniform \cite{TorquatoReview}. This feature is shown in Fig.\,(\ref{Fig1})\,(b), where we found that the density fluctuations of 3D-SHDS with $\chi\geq20\%$ scale proportionally to $R^{2.03\pm0.25}$. On the contrary, for lower $\chi$ values, $\mu_2$ scales as $\langle N_R\rangle_e$  (orange line), which is proportional to $R^3$  (see also Fig.\,(\ref{FigA1})\,(f)). 

In hyperuniform systems, the total correlation function $h(\textbf{r})=g_2(\textbf{r})-1$, which is zero when there are no spatial correlations, must become negative for some $\textbf{r}$ values \cite{LucaBibbiaBook,SgrignuoliPeenwheel}. A general approach to identify regions of negative structural correlations is based on the study of information contained in high-order correlation functions. This study can be performed by considering the moments of the number of points within a spherical window of radius R, i.e., $\mu_j=\langle(N_R-\langle N_R\rangle^j\rangle)$ \cite{SgrignuoliSubrandom,TorquatoHOC}. Specifically, the high-order correlation function $\gamma_2$, also named kurtosis, is defined through the three-point (i.e., $\mu_3$) and fourth-point (i.e., $\mu_4$) moments as $\mu_4\mu_3^{-2}-3$ \cite{Bohigas}. Figure\,(\ref{Fig1})\,(c) displays the behavior of $\langle\gamma_2\rangle_e$ as a function of $R/\langle \overline{d_1}\rangle_e$. 3D-SHDS with $\chi\geq20\%$ exhibit a range where $\langle\gamma_2\rangle_e$ is oscillatory and negative, indicating the presence of strong structural correlations with a repulsive behavior. On the contrary, point patterns generated with smaller $\chi$ values are well-described by the analytical trends predicted for uniform random point patterns (blue line) \cite{Mehta,SgrignuoliSubrandom}. Finally, we show the probability density function $P(\langle d_1\rangle_e/\langle\overline{d_1}\rangle_e)$ of the normalized ensemble-averaged nearest-neighbor distances in Fig.\,(\ref{Fig1})\,(d). We found that, independently of the degree of stealthiness, $P(\langle d_1\rangle_e/\langle\overline{d_1}\rangle_e)$ are well-reproduced by a Gaussian distribution, which describes the spatial arrangement of hyperuniform scatterers within a spherical window of radius $R$ theoretically \cite{TorquatoHOC}. Interestingly, the quantities $P(\langle d_1\rangle_e/\langle\overline{d_1}\rangle_e)$ become more peaked around their averaged values $\langle\overline{d}_1\rangle_e$ for $\chi$ larger than $20\%$. As recently observed in 3D deterministic hyperuniform structures generated from subrandom sequences, the probability of finding particles with a normalized separation larger than 0.5 facilitates light localization \cite{SgrignuoliSubrandom}. In turn, this structural feature reduces the excitation of proximity resonances (i.e., dark sub-radiant modes spatially extended over just a few scatterers \cite{Heller}) and promote the formation of spectral gaps due to multiple wave interference \cite{SgrignuoliSubrandom,Sgrignuoli2019}. This is so because the fraction of scatterers that are strongly coupled by the dipole-dipole near-field interaction, which scales with their relative distance as $1/r_{ij}^3$, is drastically reduced. 

Mathematically, hyperuniformity is defined in the thermodynamic limit (i.e., for infinite structures) \cite{TorquatoReview}. However, samples that are generated experimentally \cite{Muller,Aubry,Gorsky} or computationally \cite{Torquato,TorquatoReview,SgrignuoliSubrandom} have necessarily finite size. Recently, quantitative criteria have been established to ascertain the crossover distance of the hyperuniform/non-hyperuniform regimes in finite systems \cite{torquato2021strScaling}. According to the analysis provided in this section and in the Appendices, the investigated structures are sufficiently large to manifest robust stealthy hyperuniform behavior \cite{TorquatoReview}. Consequently, the proposed analysis enables the general  understanding of the physical mechanisms behind the light transport in the considered correlated disordered media. In particular, this article is concerned with the systematic study of optical transport and localization phenomena depending on the type and degree of structural correlations achieved in finite-size 3D media. In order to address this complex problem for the investigated 3D-SHDS we introduce in the next section the multiple scattering framework within the Green's matrix spectral method \cite{RusekPRE,Lagendijk}.

\section{Scattering and localization properties of 3D-SHDS}\label{Transport}
The $3N\times3N$ Green's matrix composed of the elements 
\begin{eqnarray}\label{GreenOur}
\begin{aligned}
G_{ij}=&i\delta_{ij}+\frac{3i}{2}\left(1-\delta_{ij}\right)\frac{e^{ik_0r_{ij}}}{ik_0r_{ij}}\Biggl\{\Bigl[\bold{U}-\hat{\bold{r}}_{ij}\hat{\bold{r}}_{ij}\Bigr]\\
&- \Bigl(\bold{U}-3\hat{\bold{r}}_{ij}\hat{\bold{r}}_{ij}\Bigr)\left[\frac{1}{(k_0r_{ij})^2}+\frac{1}{ik_0r_{ij}}\right]\Biggr\}
\end{aligned}
\end{eqnarray}
describes the electromagnetic vector coupling among $N$ point scatterers, spatially arranged to form the different 3D-SHDS investigated in this work. Here, $k_0$ is the wavevector of light, the integer indexes $i, j \in 1,\cdots,N$ label the different particles, $\textbf{U}$ is the 3$\times$3 identity matrix, $\hat{\bold{r}}_{ij}$ is the unit vector position from the $i$-th and $j$-th scatter while $r_{ij}$ identifies its magnitude. The complex eigenvalues $\Lambda_n$ ($n\in$ 1, 2, $\cdots$ 3N) of the matrix (\ref{GreenOur}) have a physical interpretation in scattering theory, where they correspond to the scattering resonances of the system. Specifically, their real parts are equal to the normalized detuned frequencies $(\omega_0-\omega_n)$, while their imaginary parts are the decay rate $\Gamma_n$ of the scattering resonances, both normalized by the decay rate $\Gamma_0$ of a bare dipole in free-space \cite{RusekPRE,Lagendijk}. 

\begin{figure*}[t!]
\centering
\includegraphics[width=\textwidth]{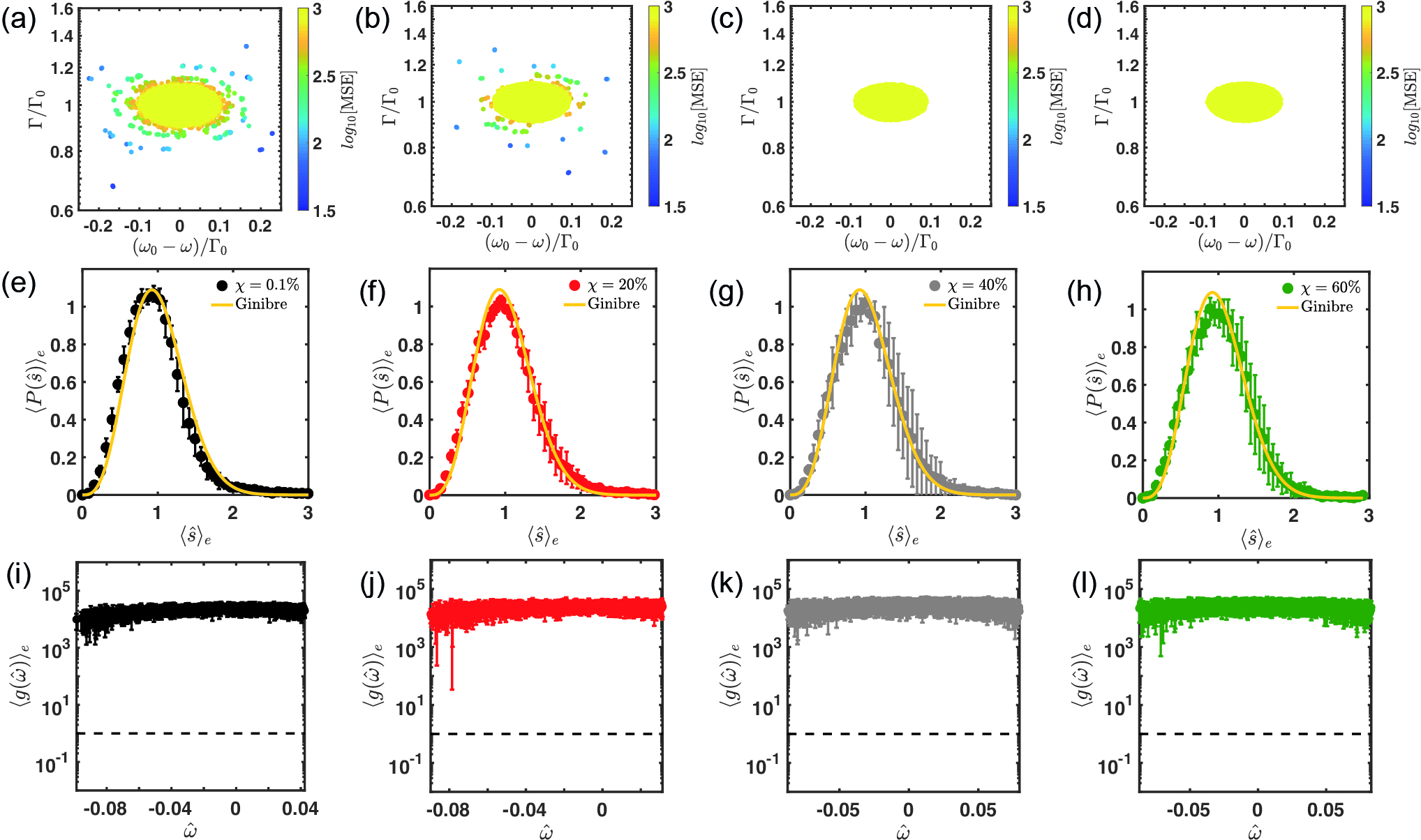}
\caption{Spectral properties of 3D-SHDS when $\rho/k_0^3=10^{-6}$. Panels (a-d) show the eigenvalues of the Green's matrix associated with the structures discussed in Fig.\,(\ref{Fig1}) color-coded according to the logarithmic of the MSE parameter. Panels (e-h) display the ensemble-averaged probability distribution functions of the level spacing statistic of the Green's eigenvalues reported on top of each panel. The prediction from Ginibre's ensemble of random matrices is reported with the yellow lines for comparison. Panels (i-l) show the corresponding Thouless numbers as a function of the frequency $\hat{\omega}$, ensemble-averaged over ten different realizations. The dashed-black lines identify the threshold of the diffusion-localization transition. The error bars are the statistical errors associated with the average ensemble operation.}
\label{Fig2}
\end{figure*}

The predictions of Green's matrix spectral method captures the physics of the multiple scattering problem in the limit of small electric dipole scatterers \cite{SkipetrovPRL,SgrignuoliSubrandom,Sgrignuoli2019,SgrignuoliCL,RusekPRE}. This approximation is valid for particles with a small size parameter $x=ka<<1$ ($k$ is the wavenumber and $a$ is the particle radius) \cite{Lagendijk, RusekPRA, DalNegroElliptic}. This approach has been extensively utilized to study the scattering properties of large-scale aperiodic media, providing access to robust transport and spectral information that cannot otherwise be accessed using traditional numerical methods such as Finite Difference Time Domain (FDTD) or Finite Elements (FEM) \cite{dal2016structural,wang2018spectral,Sgrignuoli2019,DalNegroElliptic}. The main feature of this methodology is that it allows one to treat multiple light scattering exactly in a scalable and computationally efficient way, abstracting from the material properties and sizes of the individual particles that are separately captured by a retarded polarizability \cite{Lagendijk, DalNegroElliptic}. Moreover, the Green's matrix spectral method can be extended to include external magnetic \cite{Skipetrov2015,Skipetrov2018} or electric \cite{Skipetrov2019} fields and higher-order multipolar resonances \cite{DalNegroElliptic}, which are outside the scope of the present work. Finally, it is worth noting that the non-Hermitian Green matrix (\ref{GreenOur}) is also an excellent tool to describe the behavior of atomic clouds. In particular, cold atoms spatially arranged in aperiodic atomic lattices \cite{Roati} are a suitable alternative to dielectric materials to demonstrate vector wave localization in 3D environments. Recently, quantum-gas microscopes \cite{Kuhr} enabled the engineering of one \cite{Endres}, two \cite{BarredoScience}, and even three-dimensional \cite{Greiner,WangCold,Nelson,Barredo} optical potentials with arbitrary shape while keeping single-atom control to simulate models from condensed matter physics in highly controlled environments \cite{Bloch}. Therefore, 3D optical atomic lattices based on hyperuniform disordered potentials with a tunable degree of long-range spatial correlations are also suitable platforms to experimentally demonstrate the results of this paper.

\subsection{Low Optical Scattering Regime}
In order to unveil the physical mechanisms behind light propagation in correlated disordered media, we have analyzed the scattering resonances, the level spacing statistics, and the Thouless number of representative 3D-SHDS in the low and large optical density regime. The optical density is defined as $\rho/k_0^3$, where $\rho$ denotes the dipole volume density $N/V$.
At low optical density $\rho/k_0^3=10^{-6}$, we found that all the investigated structures are in the diffusive regime, independently on the value of $\chi$. The complex distributions of $\Lambda_n$, color-coded in Fig.\,(\ref{Fig2})\,(a-d) according to the $\log_{10}$ values of the Modal Spatial Extent (MSE) \cite{SgrignuoliACS}, do not indicate the presence of long-lived scattering resonances with $\Gamma_n\ll\Gamma_0$. The MSE quantifies the spatial extension of a given scattering resonance $\Psi_i$ of an arbitrary scattering system and it is defined as \cite{SgrignuoliACS}:
\begin{equation}
\text{MSE}=\left(\displaystyle\sum\limits_{i=1}^{3N} \left|\Psi_i\right|^2\right)^2\Big/\displaystyle\sum\limits_{i=1}^{3N}  \left|\Psi_i\right|^4
\end{equation}
where $N$ indicates the total number of scattering particles. The ensemble-averaged probability density functions of the first-neighbor level-spacing distribution of the complex eigenvalues of the Green's matrix $\langle P(\hat{s})\rangle_e$ are reported in Fig.\,(\ref{Fig2})\,(e-h). The quantity $\langle P(\hat{s})\rangle_e$ is consistently modeled by the Ginibre distribution for all the investigated $\chi$. One of the most crucial features of this statistical distribution, which is defined as \cite{Haake}
\begin{equation}\label{Ginibre}
P(\hat{s})=\frac{3^4\pi^2}{2^7}\hat{s}^3 \exp\left(-\frac{3^2\pi}{2^4}\hat{s}^2\right)
\end{equation}
with $\hat{s}$ the nearest-neighbor eigenvalue spacing $|\Delta\Lambda|$=$|\Lambda_{n+1}-\Lambda_n|$ normalized to the average spacing $\overline{|\Delta\Lambda}|$, is the so-called level repulsion behavior: $P(\hat{s})\rightarrow0$ when $\hat{s}\rightarrow0$. The level repulsion is related to the extended nature of eigenmodes due to the mutual orthogonality of eigenvectors that forbids that two extended eigenvectors are degenerate \cite{Skipetrov2015}. Our analysis based on the Ginibre distribution demonstrates that the level spacing of disorder hyperuniform arrays exhibits cubic level repulsion in the weak scattering regime. Interestingly, cubic level repulsion is a quantum signature of chaos in dissipative systems irrespective of whether their Hamiltonians obey time-reversal invariance \cite{Grobe}. Finally, the diffusive nature of 3D-SHDS in the weak scattering regime (i.e., at low optical density) is also corroborated by the behavior of the Thouless number that remains larger than unity independently on the frequency $\omega$, as shown in Fig.\,(\ref{Fig2})\,(i-l). The Thouless number $g$ as a function of $\omega$ is evaluated as:
\begin{equation}\label{Thouless}
\langle g(\hat{\omega})\rangle_e=\Big\langle\frac{\overline{\delta\omega}}{\overline{\Delta\omega}}\Big\rangle_e=\Bigg\langle\frac{(\overline{1/\Im[\Lambda_n]})^{-1}}{\overline{\Re[\Lambda_n]-\Re[\Lambda_{n-1}]}}\Bigg\rangle_e
\end{equation}
\begin{figure*}[t!]
\centering
\includegraphics[width=\textwidth]{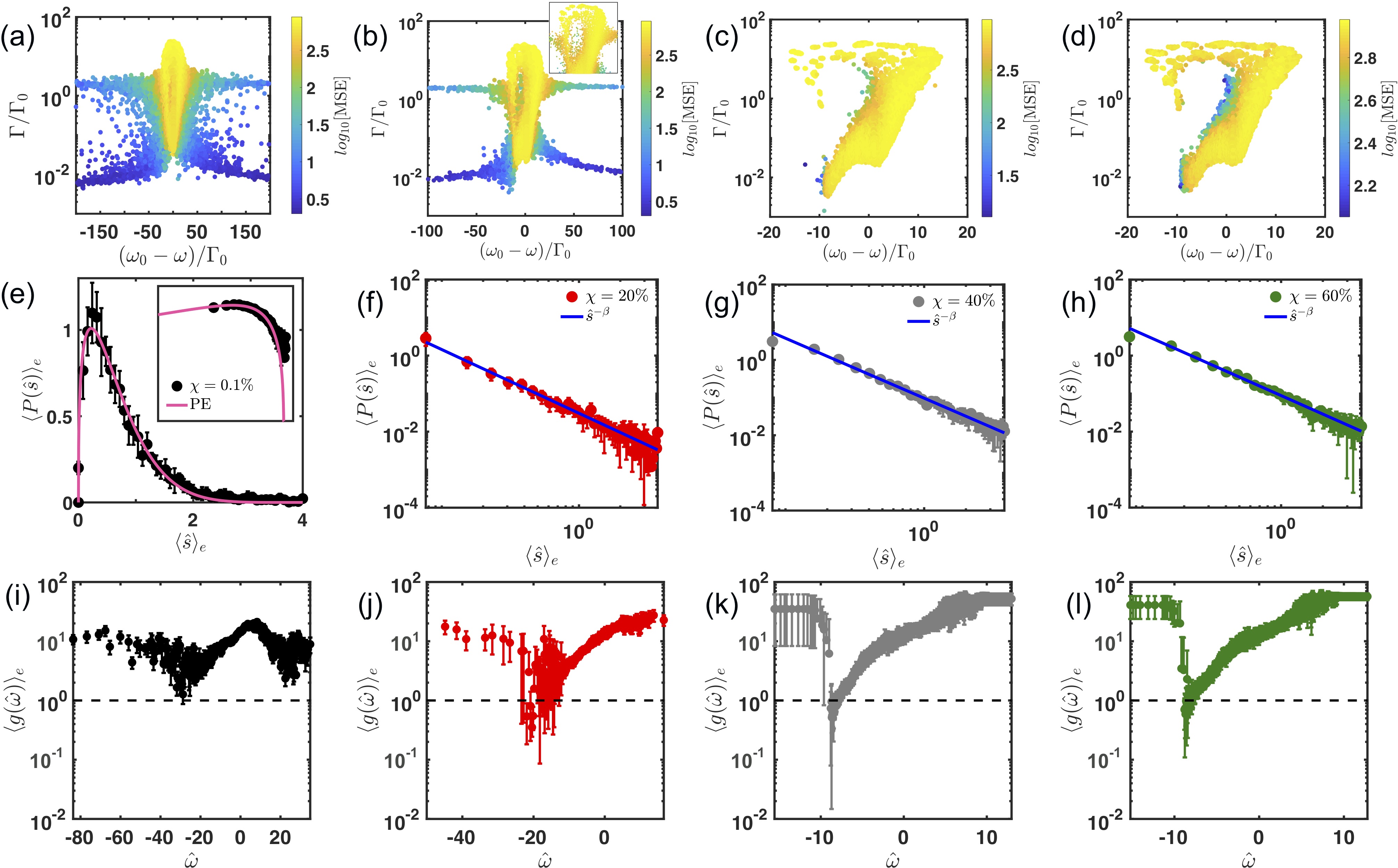}
\caption{Spectral properties of 3D-SHDS when $\rho/k_0^3=0.5$. Panels (a-d) show the eigenvalues of the Green's matrix associated with the structures introduced in Fig.\,(\ref{Fig1}) color-coded according to the logarithmic of the MSE parameter. Inset in panel (b) is an enlarged view of the spectral gap region when $\chi=20\%$. Panels (e-h) display the ensemble-averaged probability distribution functions of the level spacing statistic of the Green's eigenvalues reported on top of each panel. Inset in panel (e) shows the data in a log-log scale. While the most disordered structures characterized by $\chi=0.1\%$ show level repulsion (rose line), the other configurations display level clustering with an inverse  power-law $\hat{s}^{-\beta}$ (blue lines). Panels (i-l) show the corresponding Thouless numbers as a function of the frequency $\omega$, ensemble-averaged over ten different realizations. The dashed-black lines identify the threshold of the diffusion-localization transition. The error bars are the statistical errors associated with the average ensemble operation.}
\label{Fig3}
\end{figure*}
following the same procedure as in our previous work \cite{Sgrignuoli2019,SgrignuoliSubrandom,SgrignuoliPeenwheel}. In particular, we have sampled the real part of the Green's matrix eigenvalues in equispaced intervals and we have evaluated Eq.(\ref{Thouless}) in each frequency interval. The symbol $\overline{\{\cdots\}}$ in Eq.(\ref{Thouless}) denotes the interval averaging operation, while $\hat{\omega}$ indicates the central frequency of each interval.

\subsection{Large Optical Scattering Regime}\label{LargeRegime}
The situation is drastically different when we consider the strong scattering regime. In fact, we found that at large scattering density $\rho/k_0^3=0.5$, light transport is strongly influenced by the degree of stealthiness $\chi$, i.e., the degree of structural correlations. As it is visible in Fig.\,(\ref{Fig3}), the 3D-SHDS with $\chi=0.1\%$ shows a delocalized regime dominated by sub-radiant resonances [blue spiral arms in panel (a)], characterized by a level spacing distribution with level repulsion [panel (e)], and by a Thouless number larger than unity independently of $\omega$ [panel (i)]. Specifically, uncorrelated 3D configurations do not show any signature of light localization, i.e., the Thouless number is always larger than unity and the distribution $\langle P(\hat{s})\rangle_e$ is described by the phenomenological ensemble \cite{Sorathia,Casati}:
\begin{equation}\label{PE}
P_\beta(\hat{s})=B_1(1+B_2\beta z)^{f(\beta)}\exp\left[-\frac{1}{4}\beta z^2-\left(1-\frac{\beta}{2}\right)z\right]
\end{equation}  
where $f(\beta)=\beta^{-1}2^\beta(1-\beta/2)-0.16874$, $z=\pi\hat{s}/2$, $0\leq\beta\leq\infty$ is a fitting parameter, and $B_1$ and $B_2$ are determined by normalization conditions \cite{Casati}. Equation\,(\ref{PE}) describes the entire evolution of the eigenvalue statistics of random matrices as a function of the strength of disorder through the $\beta$ parameter. Notice that for $\beta=0$, eq.(\ref{PE}) is equal to the Poisson statistics typically associated to noninteracting, exponentially localized energy levels \cite{Haake}. For configurations with $\chi=0.1\%$, our results indicate a $\beta$ value equal to 0.5, indicating level repulsion and, therefore, the absence of localization for small values of the stealthiness parameter.

In contrast, the 3D-SHDS with larger $\chi$ display completely different features. Specifically, these structures are characterized by: (i) the formation of spectral gaps in their complex eigenvalue distributions in panels (b-d); (ii) a reduction of proximity resonances that disappear by increasing $\chi$  [see panels (b-d)]; (iii) a level spacing distribution $\langle P(\hat{s})\rangle_e$ that feature level clustering in panels (f-g) and, finally, a Thouless number that becomes smaller than unity depending on frequency $\hat{\omega}$ in panels (j-l). Interestingly, the spatial extension of long-lived scattering resonances increases by raising the degree of short-range order of the structures, as a consequence of the more ordered geometrical nature. At the same time, the spectral gaps, visible in panels (b-d), widen by increasing $\chi$. As introduced in section \ref{Geometry}, this distinctive spectral feature is directly related to the reduction of proximity resonances, weakening near-field interactions that are detrimental to vector waves localization \cite{SkipetrovPRL,Bellando,van2021longitudinal,cobus2021transient, naraghi2016phase,SgrignuoliSubrandom}.

These results unveil a link between the structural properties, reported in Fig.\,(\ref{Fig1}), and the transport ones, analyzed in Fig.\,(\ref{Fig2}) and Fig.\,(\ref{Fig3}). Three-dimensional disordered structures with suppressed large-scale density fluctuations display a transition from a diffusive to a weak localization regime because $g$ drops below unity \cite{LagendijkToday,Froufe,Aubry,Sgrignuoli2019,SkipetrovPRL,Skipetrov2020,SgrignuoliSubrandom,Skipetrov2018} and the level spacing switches from a cubic-type level repulsion to level clustering \cite{SgrignuoliSubrandom,Skipetrov2015,Shklovskii,Casati}. The discovered localization mechanism is different from the Anderson localization transition that occurs in random systems with Gaussian statistics, which are homogeneous and isotropic disorder media supporting exponentially localized modes (i.e., characterized by a level spacing distribution with a Poisson statistics). On the contrary, as visible in Figs.\,(\ref{Fig3})\,(f), (g), and (h), the level spacing distributions of 3D-SHDS are well-reproduced by the inverse power-law scaling $\langle P(\hat{s})\rangle_e\sim\hat{s}^{-\gamma}$, with values of the exponent $\gamma$ equal to $1.68\pm0.21$, $1.59\pm0.10$, and $1.62\pm0.12$, respectively. Level clustering with power-law scaling indicates anomalous transport within a sub-diffusive regime in which the width of a wave packet $\sigma^2$ increases in time with power-law scaling $t^{2\nu}$ and $\nu\in[0,1]$, as shown in references \cite{SgrignuoliSubrandom,Cvitanovic,Geisel,SebbahAnomalous}. Interestingly, sub-diffusive transport has been theoretically predicted and experimentally observed in 3D random media and associated with the occurrence of recurrent scattering loops in the transport of waves \cite{cobus2021transient,naraghi2016phase}. The anomalous exponent $\nu$ is related to the parameter $\gamma$ through the relation $\nu=(\gamma-1)/d$, where $d$ is the system dimensionality \cite{Cvitanovic,Geisel,SebbahAnomalous}. By substituting the $\gamma$ values extrapolated by using a least-square method, we find that the exponent $\nu$ is equal to $0.23\pm0.07$, $0.19\pm0.03$, and $0.21\pm0.04$ for 3D-SHDS with a degree of stealthiness equal to $20\%$, $40\%$, and $60\%$, respectively. The fact that $\nu$ is lower than 0.5 suggests that the propagation of wave packets throughout the investigated structures is sub-diffusive, indicating a significant wave interference correction to classical diffusion and a transition to the weak localization regime \cite{John}. Our findings establish a clear transition from a diffusive to a weak localization sub-diffusive regime for vector waves that propagate in finite-size three-dimensional stealthy hyperuniform random media. 

\begin{figure}[b!]
\centering
\includegraphics[width=\columnwidth]{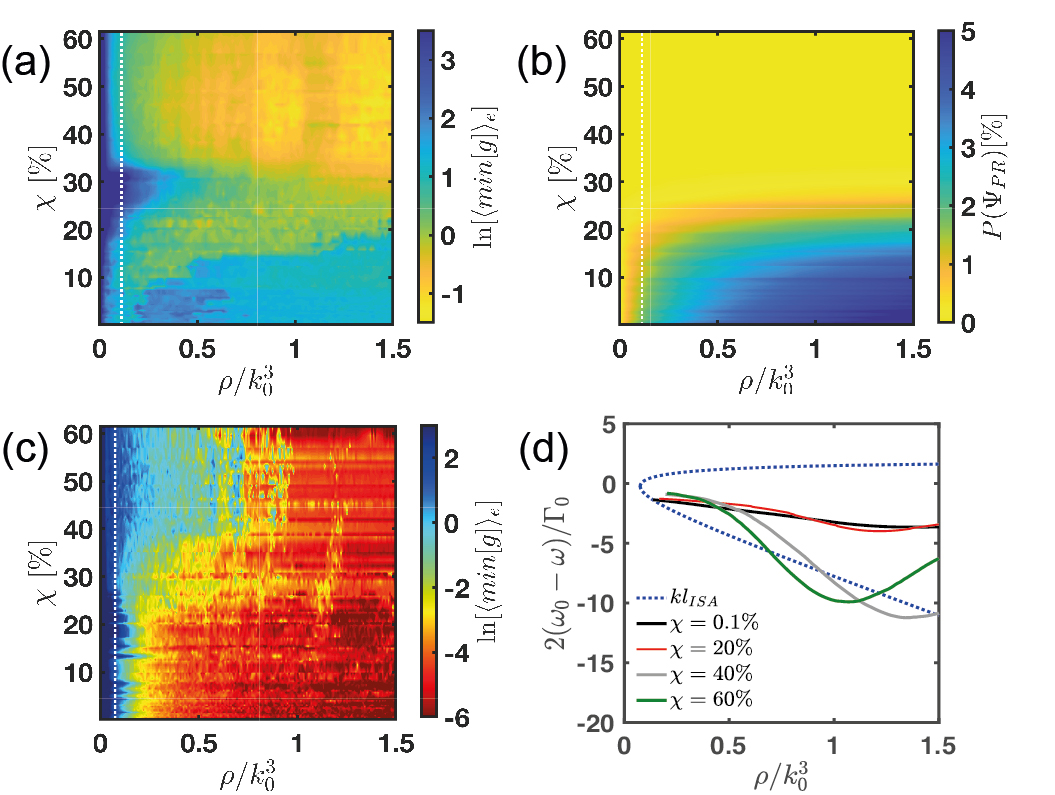}
\caption{Structural correlations and wave transport properties of 3D-SHDS. Panel (a) shows the ($\chi;\rho/k_0^3$) transport phase diagram when the vector nature of light is taken into account. Specifically, the minimum Thouless number, ensemble-averaged over ten different disorder realizations for each configurations, is reported as a function of $\rho/k_0^3$ and for several degrees of stealthiness $\chi$. The probability of the formation of proximity resonances as a function of $\rho/k_0^3$ for different $\chi$ values is reported in (b). The white dotted line in panels (a) and (b) predicts the minimum density $(\rho/k_0^3)_{c}$ needed to reach localization in the ISA model. Panels (c) and (d) summarize the transport properties by employ the scalar approximation. Panel (c) shows the ($\chi;\rho/k_0^3$) transport phase diagram, while panel (d) compares the frequency values corresponding to the minimum value of Thouless number of selected $\chi$ structures with respect to the predictions of the Ioffe-Regel criterion in the independent scattering approximation for scalar waves (blue-dashed lines).}
\label{Fig4}
\end{figure}
\subsection{Transport Phase Diagram}\label{PhaseDiagram}
An important aspect of our work is the classification of different transport regimes in 3D disordered photonic materials based on their structural correlations. We can now address this problem by generating $75$ different 3D structures characterized by a $\chi$ parameter in-between uncorrelated (i.e., $0.1\%$) and crystalline (i.e., $60\%$) regimes and solving the electromagnetic multiple scattering problem using the Green's matrix method. An overview of the structural properties of the generated arrays is presented in Appendix \ref{Appendice}. These results confirm the stealthy hyperuniform character of the investigated structures and demonstrate that relevant structural correlation effects start to occur when $\chi$ is larger than $10\%$.

Figure\,(\ref{Fig4})\,(a) displays the minimum value of the Thouless number, ensemble-averaged to different disorder realizations, as a function of $\rho/k_0^3$ and $\chi$. This transport phase diagram clearly shows that localization (i.e., negative $\ln[\langle\min[g]\rangle_e]$ values) strongly depends on the structural correlation parameter $\chi$. Interestingly, we discover a direct connection between these findings and the probability of the formation of proximity resonances $P(\Psi_{PR})$ as a function $\rho/k_0^3$ for different degrees of stealthiness, as reported in Fig.\,(\ref{Fig4})\,(b). $P(\Psi_{PR})$ is evaluated as the ratio between the sum of all the proximity resonances, identified by the condition MSE=2, over the number of all the scattering resonances of the system. Since proximity resonances arise when few identical scatterers are placed close together, these states are dominated by dipole-dipole near-field interactions \cite{Heller, SkipetrovPRL}. Fig.\,(\ref{Fig4})\,(b) shows a strong reduction in the number of proximity resonances forming in 3D-SHDS as a function of $\chi$, reflecting a decreasing of near-field effects. In the investigated structures, weak localization occurs when $\chi>10\%$. Specifically, our results suggest that this regime takes place when the near-field mediated dipole-dipole interactions are sufficiently reduced by the degree of hyperuniform stealthiness of the systems. Moreover, such localization regime cannot be observed unless a specific value of the optical density is reached (i.e., $\rho/k_0^3<0.1$). This fact is evident by comparing Fig.\,(\ref{Fig4})\,(a) with Fig.\,(\ref{Fig4})\,(b), despite the low probability of proximity resonances in the systems. Even though the discovered transition into the weak localization regime is different from Anderson light localization, it can be qualitatively understood based on the Ioffe-Regel (IR) criterion $k\ell=1$ evaluated in the independent scattering approximation (ISA) \cite{Sheng}. Here, $k=2\pi/\lambda$ is the wave number and $\ell$ is the transport mean free path that is equal to $1/\rho\sigma_{ext}$ in the ISA approximation \cite{Lagendijk}. Under the conditions of this study, the extinction cross section $\sigma_{ext}$ coincides with the scattering cross section $\sigma_{sca}$ \cite{Pinheiro2008}. The IR criterion provides a critical value of the optical density $(\rho/k_0^3)_c$ equals to $(\sqrt{5}-2)/2\pi$ \cite{Lagendijk} for the occurrence of a DLT [dotted white line in panels (a) and (b)]. The IR criterion also qualitatively explains why the investigated 3D-SHDS are in the diffusive regime for values of the optical density smaller than $(\rho/k_0^3)_c$, although $P(\Psi_{PR})$ is almost zero in this spectral range. Interestingly, the data in panel (a) show a region in the transport phase diagram where $\ln[\langle\min g\rangle_e]>0$ for values of optical densities larger than $(\rho/k_0^3)_c$ and for $20\%<\chi<35\%$. This indicates the presence of a spectral region where  transport occurs through the system despite a simple estimate based on the IR criterion and the strongly reduced probability of proximity resonances would suggest localization instead. We note that the existence of a transparency range in 3D-SHDS whit specific $\chi$ parameters compatible with the data reported in panel (a) was also recently predicted in the strong-contrast approximation of the effective dynamic dielectric constant using non-local homogenization theory \cite{torquato2021nonlocal}). 

Interestingly, the transport behavior of 3D-SHDS is completely different if we neglect the near-field mediated dipole-dipole interactions. To neglect this term, we have considered the scalar approximation of the matrix\, (\ref{GreenOur}), defined as follows \cite{SkipetrovPRL}:
\begin{equation}
\tilde{G}_{ij}=i\delta_{ij}+(1-\delta_{ij})\frac{e^{ik_0r_{ij}}}{k_0r_{ij}}
\end{equation}
Figure\,(\ref{Fig4})\,(c) displays the phase transport diagram of the minimum value of the Thouless number, ensemble-averaged over different disorder realizations, as a function of $(\chi;\rho/k_0^3)$. As expected \cite{SkipetrovPRL}, localization now takes place, independently of the value of $\chi$, when the minimum density $(\rho/k_0^3)_c$ predicted by the IR criterion for scalar waves and within the ISA approximation has been reached \cite{SkipetrovIoeffe}. We remark that for scalar waves, the IR criterion predicts a smaller $(\rho/k_0^3)_c$ threshold than for vector waves \cite{Lagendijk, SkipetrovIoeffe,Skipetrov2018}, as also shown in Fig.\,(\ref{Fig4}) by comparing the white dotted lines in panel \,(c) with respect to the one in panel (a). Moreover, it is evident that localization effects are much stronger (i.e. lower values of Thouless number) in the scalar than the vector case, which shows instead a smoother transition, as discussed in the previous section\,\ref{LargeRegime}. This fact is not surprising considering that the near-field and intermediate-field mediated dipole-dipole interactions are absent \cite{Sgrignuoli2019}. However, it is possible to identify a region of the transport phase diagram where the investigated systems are also transparent within the scalar approximation. This region starts for $\chi$ values larger than $20\%$ and for optical densities larger than $(\rho/k_0^3)_c$. This transport area increases with the degree of stealthiness and it extends up to $\rho/k_0^3<0.4$ for the largest $\chi$ value that we have considered in this work (i.e., the blue/cyan region visible in panel\,(c) that increases with $\chi$). Interestingly, this result is similar to what was observed in the light transport properties of scalar waves (i.e., TM-polarized waves) in two-dimensional disordered stealthy hyperuniform structures, where a transparency region appears for small frequencies and only for the structures with $\chi$ values larger than $20\%$ \cite{Froufe}. Moreover, the area of this transparent region also increases with the degree of stealthiness in two-dimensional arrays \cite{Froufe}. Finally, the IR criterion for scalar waves enables the identification of a spectral region where all the states are localized, i.e., $k\ell=1$ \cite{SkipetrovIoeffe}. Because both $k$ and $\ell$ depend on the frequency $\omega$, the IR criterion defines a spectral region that separates the frequencies for which the transport is diffusive from the ones for which light is localized \cite{SkipetrovIoeffe}. This localization region is contained within the blue dotted line shown in Fig.(\ref{Fig4})\,(d). 
Fig.\,\ref{Fig4}\,(d) compares the frequency values corresponding to the minimum value of Thouless number of selected $\chi$ structures as a function of $\rho/k_0^3$ with respect to the predictions of the IR criterion. Specifically, the black, red, grey, and green curves refer to structures with $\chi$ equal to 0.1\%, 20\%, 40\% and 60\%, respectively. Interestingly, our data fall within the ISA predictions for the less correlated configurations with $\chi<20\%$. On the other hand, this theory fails, as expected, to describe more correlated 3D-SHDS, underlining the necessity of deeper theoretical investigations beyond the scope of the present work.

\begin{figure*}[t!]
\begin{center}
\includegraphics[width=\textwidth]{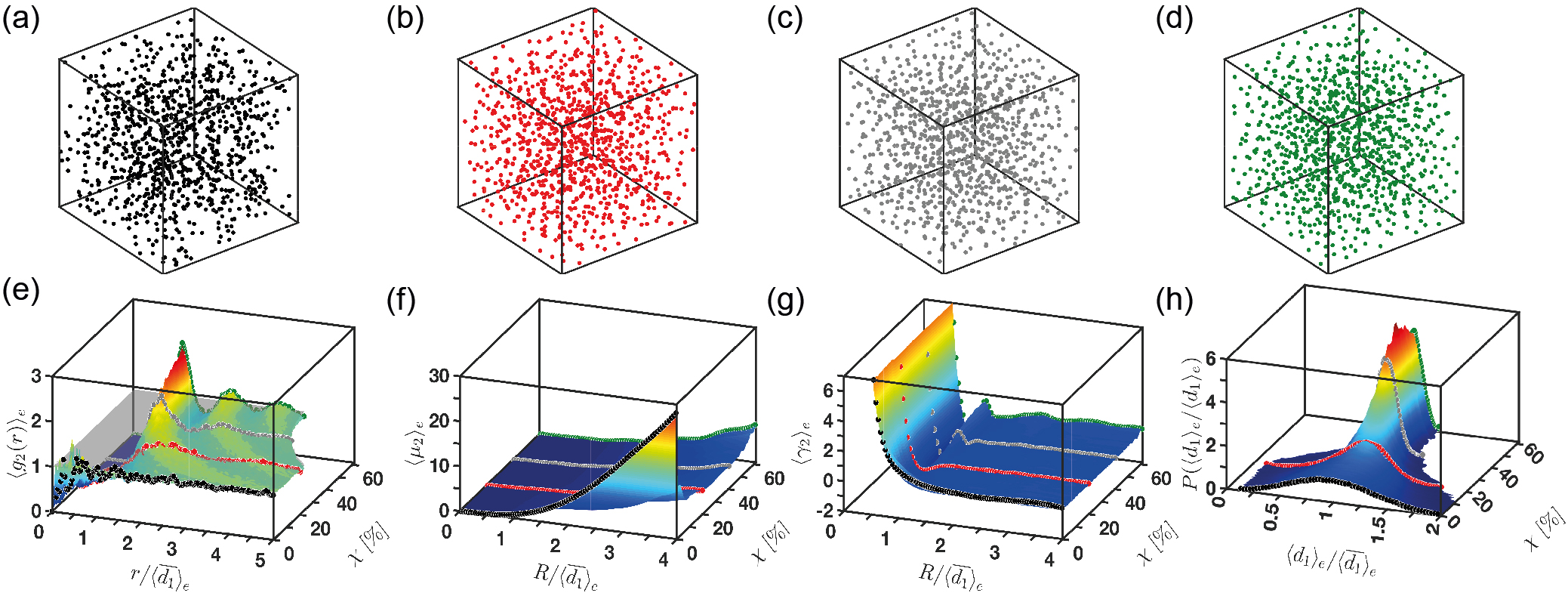}
\caption{(a) Representative stealthy hyperuniform point patterns with $\chi$ equal to $0.1\%$ (black), $20\%$ (red), $40\%$ (grey), and $60\%$ (green). Overview of the structural properties of seventy-five different 3D-SHDS. In particular, panels (e) to (h) display, the ensemble-averaged two-point correlation function $\langle g_2(r)\rangle_e$, the scaling of the number of variance $\langle\mu_2\rangle_e$ and the scaling of the $\langle\gamma_2\rangle_e$ function within a spherical observation window of radius $R$, and the probability density functions of the ensemble-averaged nearest spacing distances $\langle d_1\rangle_e$ normalized by the the ensemble-averaged of its mean value $\langle\overline{d_1}\rangle_e$. For each $\chi$ value, ten different disordered realizations are considered. Markers identify the structural properties of the representative 3D-SHDS shown in panels (a-d). Gray-shaded area in panel (e) identifies the surface $g_2(r,\chi)=1$, i.e., the uncorrelated region.}
\label{FigA1}
\end{center}
\end{figure*}


\section{Conclusions}
In conclusion, we have systematically investigated the properties of electromagnetic waves scattered by 3D stealthy disordered hyperuniform structures systematically. In particular, we have considered a large number of 3D-SHDS with a broad range of structural features in-between  crystalline and uncorrelated random materials. We have established a transport phase diagram based on the structural correlation and scattering strength properties by solving the electromagnetic multiple-scattering problem using Green's matrix spectral theory. The minimum value of the Thouless number as a function of $(\chi;\rho/k_0^3)$ clearly shows that weak localization critically depends on the structural correlation properties of the investigated systems. Moreover, we have revealed a direct connection between the mechanism of light localization and the probability of proximity resonance formation. Reducing this probability promotes the formation of spectral gaps, facilitating the onset of the discovered transition. Our numerical results demonstrate that the engineering of structural correlations is the cornerstone to understand how electromagnetic waves propagate in 3D stealthy hyperuniform environments. The present work introduces a suitable platform to experimentally test the predictions of mesoscopic physics by studying the transport of light inside 3D hyperuniform scattering environments \cite{Leonetti}. Beyond the fundamental interest in understanding the light transport properties thought 3D-SHDS, our findings suggest novel photonic media with enhanced light-matter coupling for quantum simulation protocols that are more resilient to decoherence effects \cite{weiss2017quantum,Barredo}.    
\begin{figure}[b!]
\centering
\includegraphics[width=\columnwidth]{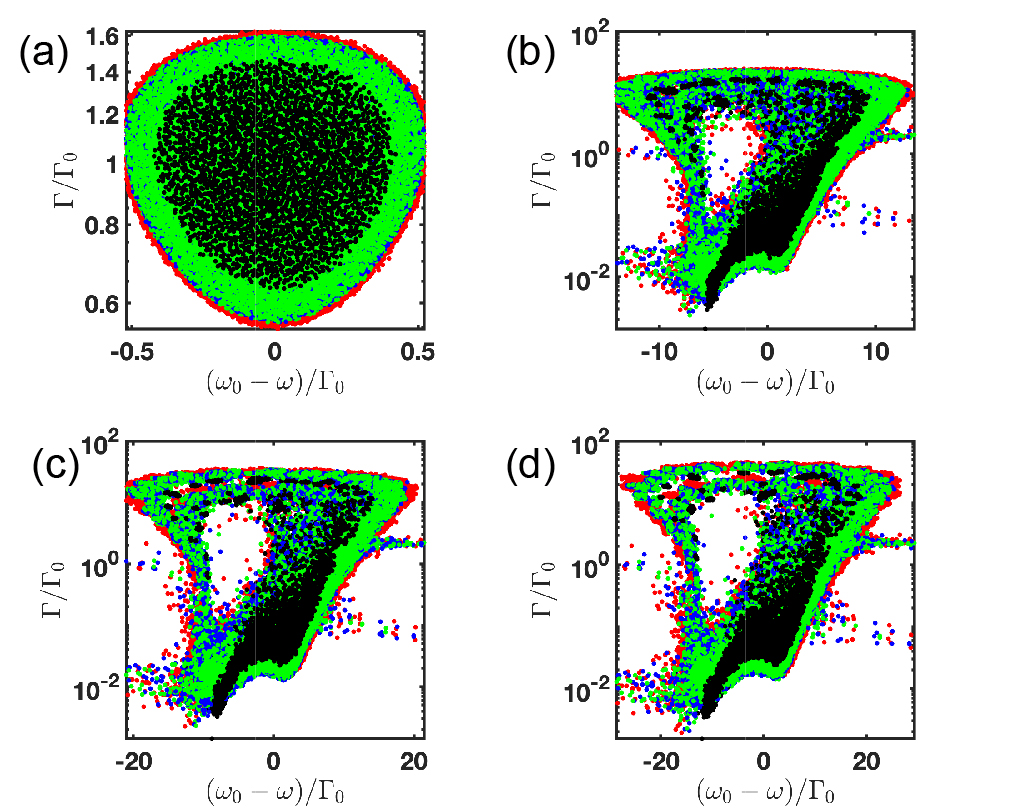}
\caption{Panels (a-d) show the Green's matrix eigenvalue distributions for different optical densities, i.e., $\rho/k_0^3$ equal to $10^{-4}$, $0.3$, $0.5$, and $0.7$, respectively. Black markers refer to systems with 1000 particles, while the red, green, and blue points refer to structures characterized by 3000 scatterers but generated with a tolerance value equal to $10^{-12}$, $10^{-15}$, and $10^{-20}$ during the point pattern optimization process. These distributions are generated by 3D-SHDS with $\chi=40\%$.}
\label{FigA2}
\end{figure}
\appendix
\section{Structural characterization of 3D-SHDS}\label{Appendice}
We have considered $75$ different 3D-SHDS characterized by different degrees of stealthiness to address the connection between structural correlation and wave transport. Figure\,(\ref{FigA1}) shows an overview of their structural properties, ensemble-averaged over ten different realizations for each $\chi$ value (i.e., a total of seven hundred fifty structures are considered). In particular, panels (a-d) show one thousand scattering elements with different degrees of stealthiness $\chi$ equal to $0.1\%$ [black points], $20\%$ [red points], $40\%$ [grey points], and $60\%$ [green points]. These structures are the ones discussed in Figs.\,(\ref{Fig1}), \,(\ref{Fig2}), and \,(\ref{Fig3}). Instead, panels (e), (f), (g), and (h) display the quantities $\langle g_2(r)\rangle_e$, $\langle\mu_2\rangle_e$, $\langle\gamma_2\rangle_e$, and  $P(\langle d_1\rangle_e/\langle\overline{d}_1\rangle_e)$. These quantities are useful to  characterize the hyperuniformity of the generated structures \cite{torquato2021strScaling,SgrignuoliSubrandom}. In particular, they allow us to identify the $\chi$ value after which the investigated structures possess long-range structural correlations. It is evident from panel (e) that the generated structures interpolate in a tunable fashion between uncorrelated random media and crystalline materials. The $r/\langle r\rangle_e$ dependence of the ensemble-averaged two-point correlation function shows clear peaks for $\chi$ values larger than $10\%$ that become more pronounced as the value of the degree of stealthiness increases. Specifically, the evolution of three distinctive peaks is visible in panel (e) by increasing $\chi$. The appearance of peaks in the two-point-correlation function indicates the transition from the fully disordered [i.e., see the grey-area in panel (e)] to the crystalline regime \cite{TorquatoReview}. Moreover, Fig.\,(\ref{FigA1})\,(e) also displays an exclusion region that increases with $\chi$ where the two-point correlation function vanishes  (i.e., the blue area nearby the origin). Consistently, the variance of the number of points contained within a spherical window of radius $R$ [see panel\,(f)] shows a gradual and a smooth change of its scaling law as a function of $\chi$. We observe a change from a volumetric to a surface growth when $\chi$ is larger than $10\%$, demonstrating the hyperuniform nature of the generated structures \cite{TorquatoReview}. As discussed in section\,\ref{Geometry}, negative values in the kurtosis function are associated to the presence of hyperuniformity. Figure\,(\ref{FigA1})\,(g) displays the formation of a region around $R/\langle\overline{d}_1\rangle_e=0.8$ that starts from $\chi$ equal to $10\%$ where $\langle\gamma_2\rangle_e$ becomes negative, indicating the presence of strong structural correlations with repulsive behavior \cite{TorquatoHOC,SgrignuoliSubrandom}. Finally, panel\,(h) shows the probability density function of the normalized ensemble-averaged nearest-neighbor distances $\langle d_1\rangle_e/\langle\overline{d}_1\rangle_e$ for different $\chi$ values. By increasing $\chi$, these Gaussian distributions become more peaked around their averaged values. As discussed in section\,\ref{Transport}, this structural feature is associated to the reduction of the proximity resonances (i.e., reduction of the near-field mediated dipole-dipole interactions) and facilitates the occurrence of localization effects when the vector nature of light is taken into account.  

\section{Stability of the spectral properties}\label{Appendice2}
\begin{figure}[b!]
\centering
\includegraphics[width=\columnwidth]{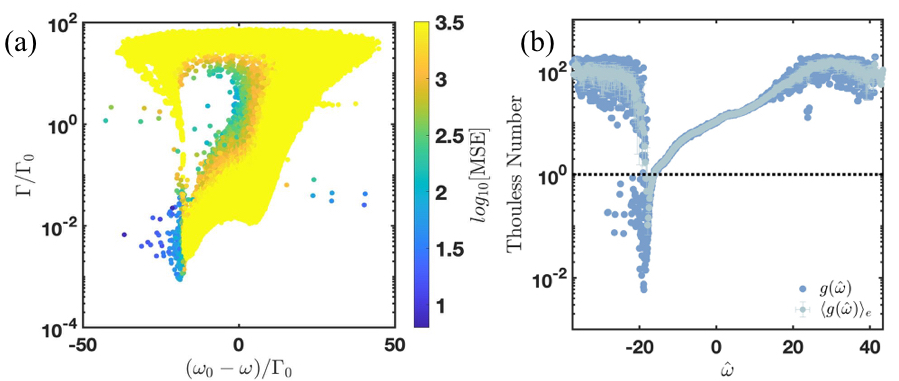}
\caption{Panel\,(a) shows the eigenvalues of the Green's matrix color-coded according to the logarithmic of the MSE parameter. Panel (b) shows the Thouless number as a function of the frequency $\hat{\omega}$ corresponding to 15 different disorder realizations (dark-pastel blue markers) and their corresponding ensemble average values (pastel blue markers).}
\label{FigA3}
\end{figure}

In this appendix, we discuss the stability of the spectral properties of the investigated arrays with respect to both the system size and the tolerance value used during the point pattern optimization process. The outcomes of this study are summarized in Fig.\,(\ref{FigA2})\,(a-d) for different optical density values, as specified in the caption, and in Fig.\,(\ref{FigA3})\,(a-b). Specifically, Fig.\,(\ref{FigA2})\,(a-d) display the Green's matrix eigenvalue distributions obtained by diagonalizing the matrix (\ref{GreenOur}) relative to ten different realizations of 3D-SHDS with $\chi=40\%$. The red, green, and blue points refer to structures with three thousand scatterers generated with a tolerance value equal to $10^{-12}$, $10^{-15}$, and $10^{-20}$ during the point pattern optimization process. These eigenvalue distributions almost coincide. This finding demonstrates the robustness of the spectral properties of the investigated structures with respect to this optimization parameter. Fig.6 also shows that the overall distributions of the complex eigenvalues of the matrix (\ref{GreenOur}) for N = 1000 (i.e., black markers in Fig.\,(\ref{FigA2})) already capture the spectral features of larger systems comprising of 3000 electric dipoles. In particular, these distributions are not influenced by proximity resonances, have the same shape independently of the optical density, and are characterized by spectral gaps in panels (b-d), which widen as $\rho/k_0^3$ increases. 
As demonstrated in Refs.\cite{SgrignuoliSubrandom, Sgrignuoli2019} and discussed in section \ref{LargeRegime}, the formation of these spectral gaps drastically reduces the appearance of proximity resonances and substantially attenuate the near-field mediated dipole-dipole interactions that are detrimental to the localization of light waves in uncorrelated random media \cite{SkipetrovPRL}. 

To study the spectral properties of larger systems, we have generated 15 different realizations of 3D-SHDS with $\chi=40\%$ and 8000 elements. Moreover, to test the stability of our findings with respect to different approaches utilized for engineering stealthy hyperuniform states of matter, we have generated these structures starting from "entropically" favored states very near-zero temperature using the algorithm described in Ref.\,\cite{zhang2015ground}. Figure\,\ref{FigA3}\,(a) shows their complex eigenvalue distributions when $\rho/k_0^3=0.5$. Interestingly, these results are consistent with the ones in Fig.\,\ref{Fig3}\,(c). In particular, the eigenvalue distributions of larger 3D-SHDS are characterized by a substantial reduction of proximity resonances. Moreover, much more structured spectral gaps form with more delocalized optical resonances. This fact is not surprising because the spatial extension of the extended scattering resonances scales with the system dimension. Interestingly, the width of  these spectral gaps is also consistent with the ones visible in Fig.\,\ref{Fig3}\,(c). In summary, the distributions of Fig.\,\ref{FigA3}\,(a) have the same features as the ones characterizing smaller finite 3D-SHDS with the difference of a larger portion of scattering resonances with slightly lower decay rate values forming close to the left edge of the spectral gaps. Notably, these optical modes also have smaller modal spatial extent values indicating the formation of Efimov-type few-body scattering resonances. The effect of these resonances in the interaction of vector waves with these structures is visible in Fig.\,\ref{FigA3}\,(b) that shows the Thouless number of 15 different realizations (dark-pastel blue markers) over imposed with their ensemble averaged values (pastel blue markers). These data display slightly lower Thouless numbers of the ones reported in Fig. 3(k) demonstrating the validity of the results presented in the main manuscript. The shift of the minimum position of the Thouless number is due to slightly different geometrical parameters utilized in the generation of larger structures. Therefore, this analysis demonstrates that the structures investigated in this work are sufficiently large to manifest the important effects of hyperuniformity in the transport and weak localization behavior of vector waves in stealthy three-dimensional environments.

\begin{acknowledgments}
This research was sponsored by the Army Research Laboratory and was accomplished under Cooperative Agreement Number W911NF-12-2-0023. The views and conclusions contained in this document are those of the authors and should not be interpreted as representing the official policies, either expressed or implied, of the Army Research Laboratory or the U.S. Government. The U.S. Government is authorized to reproduce and distribute reprints for Government purposes notwithstanding any copyright notation herein.
\end{acknowledgments}

%

\end{document}